\newcommand{\be}{\begin{equation}}
\newcommand{\ee}{\end{equation}}
\newcommand{\ba}{\begin{eqnarray}}
\newcommand{\ea}{\end{eqnarray}}
\newcommand{\bc}{\begin{center}}
\newcommand{\ec}{\end{center}}

\def\krk{k{\partial f_k(s, \Lambda)\over {\partial k}}}

\def\D{{\cal D}}             
\def\vp{\varphi }            

\documentclass[12pt,a4paper]{article}
\usepackage[cp850]{inputenc}
\usepackage{amssymb}
\usepackage{latexsym}
\usepackage[dvips]{epsfig}
\usepackage{pstricks,pst-node}
\renewcommand{\baselinestretch}{1.3}
\begin{document}
\begin{titlepage} 
\renewcommand{\baselinestretch}{1}

\vspace{0.1cm}

\begin{center}   

{\Large \sc Spontaneous Symmetry Breaking 
and Proper-Time Flow Equations}

\vspace{1.4cm}
{\large 
Alfio Bonanno} \\
\vspace{0.4cm}
\noindent
{\it INAF - Osservatorio Astrofisico,
Via S.Sofia 78, I-95123 Catania, Italy\\
INFN Sezione di Catania, Via S.Sofia 64, I-95123 Catania, Italy}

\vspace{0.5cm}
{\large 
Giuseppe Lacagnina} \\
\vspace{0.4cm}
\noindent
{\it Institut f\"{u}r Theoretische Physik, Universit\"{a}t Regensburg,\\
D-93040 Regensburg, Germany}

\end{center}   

\vspace*{0.6cm}
\begin{abstract}
  We discuss the phenomenon of spontaneous symmetry breaking by means
  of a class of non-perturbative renormalization group flow equations
  which employ a regulating smearing function in the proper-time
  integration. We show, both analytically and numerically, that the
  convexity property of the renormalized local potential is obtained
  by means of the integration of arbitrarily low momenta in the flow
  equation.  Hybrid Monte Carlo simulations are performed to compare
  the lattice Effective Potential with the numerical solution of the
  renormalization group flow equation. We find very good agreement
  both in the strong and in the weak coupling regime.
\end{abstract}
\end{titlepage}
%
\section{Introduction}
\renewcommand{\theequation}{1.\arabic{equation}}
\setcounter{equation}{0}
\label{3intro}
The mechanism of spontaneous symmetry breaking is one of the most
important nonperturbative phenomena in quantum field theory and
statistical mechanics. Unfortunately, due to its strong
nonperturbative character, many of its aspects are not yet well
understood mainly because of the lack of reliable computational tools
even in simple physical situations. For example, in the liquid-vapor
phase transition, the system undergoes a first-order phase transition
below the critical temperature, but the presence of long-range
fluctuations renders the mean-field description inadequate.  The
well-known Maxwell construction \cite{uno} reproduces the flat
isotherms in the two-phase region but does not provide the correct
critical exponents near the critical region.

In quantum field theory the situation is even more dramatic because the
perturbative calculation of the Effective Potential (EP) generates an
imaginary part whose physical meaning is not entirely clear.  In fact,
the EP should be a convex function of the fields by construction,
since it is the Legendre transform of the generating functional of the
connected Green's functions \cite{weinbergwu}.

An important result was obtained in \cite{ringwald} within the
framework of the Effective Average Action, for a continuous $O( N
)$-symmetric scalar theory: it was shown that in order to correctly
reproduce the convexity property of the free energy a non-trivial
saddle point (spin wave solution) must be chosen in the loop-expansion
\cite{tetrawette}. This result has opened the possibility of using the
Wilsonian RG approach in discussing non-perturbative issues of this
kind.

Quite generally the possibility of applying renormalization group (RG)
methods to the description of first order phase transition has
frequently been matter of discussion in the recent years.  Although
the question is still not completely settled, several works have shown
that the Wilsonian RG approach is a promising tool for such kind of
investigations.  In particular it has been shown \cite{parola} that
the Wegner-Houghton (WH) RG equation (a sharp cut-off momentum-space
wilsonian-type of RG flow equation) already in the simple local
potential approximation (LPA) reproduces the convexity property of the
free energy as a result of the integration of arbitrarily low momenta
without resorting to any {\it ad hoc} Maxwell construction, or saddle
point approximation. Moreover, the RG transformation is always well
defined below the critical line.

The essential ingredient of their analysis was the use of a robust and
very accurate numerical method for handling the RG equation without
resorting to polynomial truncations. 
Similar results have also been obtained with a smooth cutoff
RG flow equation, in the framework of the exact RG equation for the
effective average action \cite{rep1}.  In \cite{polob} an alternative
method which combines the RG flow equation approach with the
non-trivial saddle point expansion method has been presented.

The wilsonian RG transformation \cite{wilson}, as opposed to the more
conventional RG transformation based on the rescaling properties of
the Green's functions, preserves all the infinite number of
interactions generated in the low-energy action by ``averaging-out''
the small-scale physics defined at some high-energy scale $\Lambda$,
and for this reason is a powerful tool in order to investigate
situations where infinitely many length scales are coupled together.
Quite generally, the Wilsonian action can be defined as

\be e^{-S_{\Omega}(\Phi)}= \int \D \vp \delta (\Phi - C(\vp) )\ e^{-S (\vp)}
\label{eq:wilson} 
\ee 

where $\Phi$ is a low frequency field which is in turn a function of the
fundamental field $\varphi$ via $\Phi = C(\phi)$ and $C$ is some 
averaging operator.  If the effective theory is defined 
on a lattice $\Lambda_0$, then
$$ \delta (\Phi -C(\vp ))= \prod_{x\in \Lambda_0}
  \delta (\Phi(x)-C(\vp )(x)) $$
  
  For computational purposes, a continuous wilsonian RG transformation
  is defined in the momentum-space representation where the definition
  (\ref{eq:wilson}) becomes \be e^{-S_{k}(\Phi)}= \int \D \zeta e^{-S
    (\zeta + \Phi)}\label{wilson2} \ee $\zeta$ being the
  high-frequency part of the fundamental field with momenta $p$
  greater than some infrared cutoff $p>k$, and $\Phi$ the low-frequency
  one, with momenta $p\leq k$.  If we choose \be C(\vp)= {1 \over
    \Omega} \int_\Omega \vp \ee\label{ave} where $\Omega$ is a
  characteristic volume over which the field is averaged, in the limit
  $\Omega \to \infty$ (equivalently, vanishing infrared cutoff $k$ in
  Eq.(\ref{wilson2}) ) the blocked action gives precisely the EP,
  namely the non-derivative part of the Effective Action
  $\Gamma[\phi]$, defined as the Legendre transform of $W[J]$, the
  generator of the connected Green's functions in the presence of an
  external current $J$ \cite{const}.

In practice, the wilsonian RG transformation obtained from explicit
calculation of (\ref{wilson2}) is strongly non-linear, and only by
retaining infinitely many interactions generated by integrating out
the ``fast'' variable, the RG transformation can be well defined below
the critical line. Therefore, a numerical treatment of the full
non-linear partial differential equation is necessary even in the
simplest LPA approximation of the Wilsonian Action.

In recent years, in an attempt of calculating (\ref{wilson2})
non-perturbatively, a new type of RG flow equations has been proposed
\cite{propo} and further developed \cite{papp,pirnertutti,bohr}. It
is based on Schwinger's proper time  regulator, and it is
therefore well suited for a direct application to gauge theories.

The proper time renormalization group equation (PTRG) 
does not belong to the class of exact-RG flows which can be
formally derived from the Green's functions generator without
resorting to any truncation or approximation
\cite{litim1,litim2,litim3}. Nevertheless the flow equation has
the property of preserving the symmetries of the theory and, in
addition, it has a quite simple and manageable structure. Moreover the
PTRG flow, although it can not fully reproduce perturbative expansions
beyond the one loop order \cite{zappa1}, still does provide excellent
results when used to evaluate the critical properties such as the
critical exponents of the three dimensional scalar theories at the
non-gaussian fixed point \cite{boza,maza}. 
In particular the
determination is optimized when one takes the ``sharp'' limit of the
cutoff function on the proper time and it turns out to be much more
accurate than the one corresponding to a smoother regulator.  The PTRG
also provides excellent determinations of the energy levels of the
quantum double well \cite{zappa2}.
In fact, the use of the special type of PT regulator described in this paper
and already discussed in  \cite{boza,zappa2} realizes an localized integration
over the ``fast'' variables at a given infrared running
momentum $k$ and, for this reason, it shows an impressive numerical
convergence stability in the calculations of the critical exponents.

In this paper we shall instead be mainly concerned with the study of a
a strong non-perturbative phenomenon like the SSB, for which the
validity of PTRG equations has not been tested yet. In particular, the
relevant question we would like to address in this paper is to
analytically and numerically study the approach to the convexity for
the EP obtained by means of a class of PTRG flow equations in the
limit of vanishing infrared cutoff. A smeared type of proper-time
regulator will be used which selects a localized integration in the
``fast'' variables in the path-integral. The resulting flow equation
can be thought of as an interpolation of a ``smooth'' modification of
the LPA approximation of the Wegner-Houghton (WH) \cite{wh} ``exact''
flow equation, and the exponentially ultraviolet (UV) converging flow
discussed in \cite{litim1,maza}.  It will be shown that the PTRGs can
correctly describe the approach to convexity as $k\to 0$, and that
give results which are completely consistent with the analysis
performed in \cite{parola} for the WH equations, and in
\cite{tetrawette} for the ``exact'' evolution equation for the
effective average action.  The additional advantage is that below the
upper critical dimension the PTRG flow equation predicts the expected
discontinuity of the inverse compressibility, whereas the ``exact'' WH
equation fails to do so.

In order to strengthen our analysis, we compare the result of our
numerical investigation with the standard EP computed on a lattice by
means of a Hybrid Monte Carlo approach.  Our main conclusion is that
lattice EP and PTRG flow equation agree extremely well even for
relatively small lattices, both in the perturbative and
non-perturbative regime.

\section{Flow equations from proper-time regulators}
\renewcommand{\theequation}{2.\arabic{equation}}
\setcounter{equation}{0}
\label{3S2}

Let us now review the basic assumptions in the derivation of the PTRG
flow equation. We first calculate the wilsonian action in
(\ref{wilson2}) in the one-loop approximation as

\be
\label{eq:g1}
S^{1-loop}_k(\Phi) = -{1\over 2} \; \int_0^\infty \;{ds\over s} 
f_k (s,\Lambda) {\rm Tr} \; 
\Big( e^{-s{ \delta^2 S_\Lambda \over \delta\vp^2 }\Big |_{\vp=\Phi} }-
e^{-s{ \delta^2 S_\Lambda \over \delta\vp^2}\Big |_{\vp =0}} \Big )
\ee

where, following \cite{olesc,liaooc,liaoym}, we have introduced an heat-kernel
smooth regulator $f_k$ in the Schwinger proper time representation for
the one-loop expression and we have subtracted the UV divergent
contribution of a vanishing background field. The following prescription
apply to the choice of $f_k$:

\begin{itemize}
\item $f_k(s=0,\Lambda) =0$
in order to regularize the UV divergences as $s\to 0$
\item $f_{k=0}(s\to \infty, \Lambda) =1$ in order to leave the IR physics 
unchanged by the regulator
\item $f_{k=\Lambda}(s,\Lambda) =0$ in order 
to recover the original bare theory at the cutoff scale $k=\Lambda$.
\end{itemize}

The flow equation for the Wilsonian action $S_k$ results from
considering Eq.(\ref{eq:g1}) in the infinitesimal momentum shell $k,
k+\delta k$ and then {\it RG-improving the resulting expression}, which
amounts to substitute the bare action $S_\Lambda$ with the running one $S_\Lambda \to
S_k$ in the RHS of Eq.(\ref{eq:g1}),

\be
\label{flow} 
{ k{{\partial S_k(\Phi)}\over{\partial k}} =-{1\over 2}\int_0^{\infty}{ds\over s}\; 
\krk {\rm Tr}\Bigl(e^{-sS_k''(\Phi)}-e^{-sS_k''(0)}\Bigr).}  
\ee

where primes indicate functional derivatives with respect to $\vp$. At this
point, one should stress that although the fast momentum integration
in (\ref{eq:g1}) has been obtained by assuming a trivial saddle point
in the path-integral, the RG-improved flow equation (\ref{flow}) is
instead a functional differential flow equation which is valid for a
general average action $S_k(\Phi)$.

If we then project Eq.(\ref{flow}) onto its non-derivative part, the
anomalous dimension $\eta$ is set to zero, and the resulting flow
equation for the local potential reads \be\label{flow2} { k{{\partial
      U_k(\Phi)}\over{\partial k}} =-{ K_d\over
    2}\int_0^{\infty}{ds\over s^{1+d/2}} \krk\Bigl(e^{-U_k''(\Phi)
    s}-e^{-U_k''(0)s} \Bigr).}  \ee where
$K_d=2/(4\pi)^{d/2}\Gamma(d/2)$.  For actual calculations we shall
employ the following class of proper-time smearing functions \be
f^{n}_k(s,\Lambda) = \frac{\Gamma(n, s k^2) -\Gamma(n, s
  \Lambda^2)}{\Gamma{(n)}} \ee

where $\Gamma(a,z) = \int_0^z dt\; t^{\alpha -1} e^{-t}$ is the incomplete Gamma
function and $n$ a positive real number.  Note that $\lim_{n\to
  \infty}=f^{n}_k(n s,\Lambda) = \theta(s-k^2)-\theta(s-\Lambda^2)$.  

Since $k{\partial \Gamma(n,sk^2) / \partial k} = -2(s k^2)^{n} e^{-sk^2}$
the $s$-integration in (\ref{flow2}) can be explicitly performed,
having at last \ba\label{setI} && k{{\partial U_k(\Phi)}\over{\partial
    k}} = -\frac{K_d}{2} \; k^d \ln \Big( 1+\frac{U''_k(\Phi)}{k^2}
\Big ) \;\;\;\;\;\;\;\;\;\;\;\;\;\;\;\;\;\;\;\;\;\;\;\;\;\;
n=\frac{d}{2} \\[2mm ]\label{setII} && k{{\partial
    U_k(\Phi)}\over{\partial k}} = \frac{M_n^d}{(4\pi)^{d/2}} \; k^d
\Big( 1+ \frac{U''_k(\Phi)}{n k^2}\Big)^{d/2-n}
\;\;\;\;\;\;\;\;\;\;\;\;\;\;\;\;\;\;\;\;\; n > \frac{d}{2} \\[2mm]
&& k{{\partial U_k(\Phi)}\over{\partial k}} = \frac{1}{(4\pi)^{d/2}}
\; k^d \exp \Big (-\frac{U''_k(\Phi)}{k^2} \Big )
\;\;\;\;\;\;\;\;\;\;\;\;\;\;\;\;\;\;\;\;\;\;\; n \to \infty
\label{setIV} \ea where $M_n^d=n^{d/2}\Gamma(n-d/2)/\Gamma(n)$, and in
particular $M_\infty^d=1$.  In order to simplify the notation we have
not explicitly written in the RHS of the equations the field
independent contributions which correspond to a trivial shift of the
vacuum energy at $k=0$. Eq.(\ref{setI}) is obtained from the $n=d/2$
limit by performing the trivial rescaling $k\to \sqrt{d/2} k$ of the
infrared cutoff.  It coincides with the LPA approximation of the
``exact'' WH sharp cut-off equation \cite{liaooc}.  Eq.(\ref{setII})
converges to Eq.(\ref{setIV}) as $n\to \infty$. 
In other words, the
$n$-dependence in the cutoff interpolates between a WH type of equation and
the ``exponential''  RG equation (\ref{setIV}). 

\section{Analytical and numerical results}
\renewcommand{\theequation}{3.\arabic{equation}}
\setcounter{equation}{0}
\label{4S1}
We shall now analytically and numerically study the RG flow below the
critical line.  Eq.(\ref{flow}) is in principle an evolution equation
for a generic Wilsonian action $S_k(\Phi)$ and already in the simplest
LPA approximation described in (\ref{setI}-\ref{setIV}) it directly
exhibits the approach to convexity as $k\to 0$. In other words, the
non-perturbative features like spin waves, kinks or instantons are
already included in Eq.(\ref{flow}).  This result should not come
entirely as a surprise since we already stressed that (\ref{setI})
obtained for $n=d/2$, is the LPA approximation of the ``exact''
Wegner-Houghton equation studied in \cite{parola}. In this respect,
Eq.(\ref{setII}) and Eq.(\ref{setIV}) can be thought of as smooth-cutoff
modifications of the sharp cutoff WH equation, and it is then
important to understand their IR behaviour below the critical
temperature. In particular, the location of the polar singularity in
Eq.(\ref{setII}) in the $(\Phi,k^2)$ plane seems to be $n$-dependent:
\be n k^2+U''_k(\Phi)=0 \ee
so that in the limit $ n\to \infty$, 
it becomes an essential singularity at $k=0$ in Eq.(\ref{setIV}).  The relevant
question, is to understand how the convexity of the free energy is recovered 
in the $k\to 0$ limit in these cases. 

Our discussion will be mainly concerned with $n>d/2$, 
because the $n=d/2$ case has extensively discussed by \cite{parola} and 
the $n \to \infty$ case can be often extrapolated as a large $n$ regime of 
(\ref{setII}). (numerically one does not see any relevant difference 
between (\ref{setII}) and (\ref{setIV}) already for $n\sim 20$).

In order to study the behavior of the solution of Eq.(\ref{setII}) 
near $\Phi=0$ it is convenient to introduce the rescaled potential and field

\be\label{sca}
V_k = M_n^d/(4\pi)^{d/2} U_k \;\;\;\;\;\;\;\;\;\;\;
\phi = \sqrt{M_n^d/(4\pi)^{d/2}} \Phi.
\ee 

It is then convenient to take the second derivative of 
Eq.(\ref{setII}) with respect to $\phi$ and 
to define a new variable $W_k(\phi)$ through 
\be\label{u1}
W_k(\phi)=\Big ( 1+\frac{V''_k}{ n k^2 } \Big )^{d/2 - n }
\ee
which becomes large and positive in the broken phase near $\phi=0$ for  $k\to 0$.
In terms of this new variable Eq.(\ref{setII}) becomes 
\be\label{eq01}
W''_k
= - 2 k^{-d+2} n (1 -{W_k^{-\frac{1}{n-d/2}}} ) -\frac{n k^{-d+2}}{n-d/2}\;
{W_k^{-\frac{n+1-d/2} {n -d/2}}} \; k {\partial W_k \over \partial k}
\ee
and prime means now derivative wrt $\phi$.
In Eq.(\ref{eq01}),
when $W_k$ is large and positive, 
we can neglect terms which are suppressed 
as inverse power of $W_k$
and obtain the solution 
\be\label{u2}
W_k=  n\; k^{-d+2}(\phi^2_0-\phi^2) 
\ee
with $ | \phi | < \phi_0$, which is well behaved for any value of $k>0$,
being $\phi_0$ an integration constant. 
Since our approach breaks down at $\phi=\phi_0$, we can thus identify 
$\phi_0$ as the value of the field
at the coexistence. 

By inserting solution
(\ref{u2}) back in (\ref{u1}) and solving for $U_k''(\Phi)$ we have the
well-known behavior

\be
\label{inner1} 
U_k(\phi) = - \frac{ n k^2}{2}\phi^2 + O(k^2\phi^2) 
\ee 

which is consistent with the analysis of
\cite{ringwald,tetrawette}. 
Incidentally, we note that
(\ref{inner1}) shows that the approach to the flat bottom is slower
for larger values of $n$. 
\renewcommand{\baselinestretch}{1}
\small\normalsize
\begin{figure}[ht]
        \epsfxsize=14cm
        \epsfysize=10cm
\centerline{\epsffile{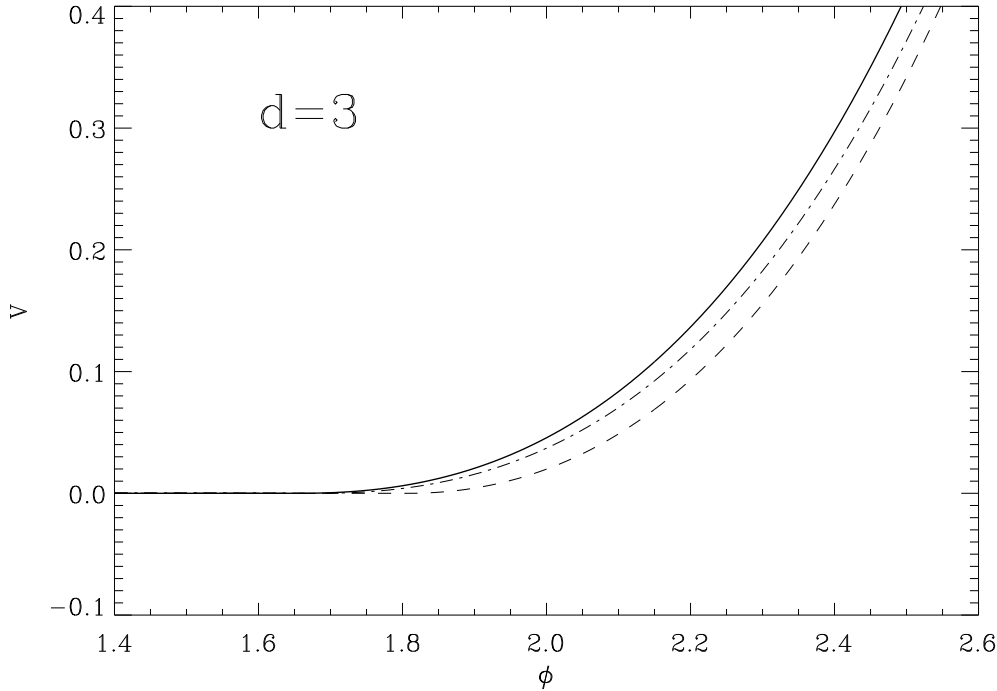}}
\caption{The blocked potential in $d=3$. 
The solid line is for $n=1.5$ (WH), the 
dashed is for $n=2$ and the dot-dashed is for $n=4$}
\label{plot1}
\end{figure}
\renewcommand{\baselinestretch}{1.5}
\small\normalsize

\renewcommand{\baselinestretch}{1}
\small\normalsize
\begin{figure}[ht]
        \epsfxsize=14cm
        \epsfysize=10cm
\centerline{\epsffile{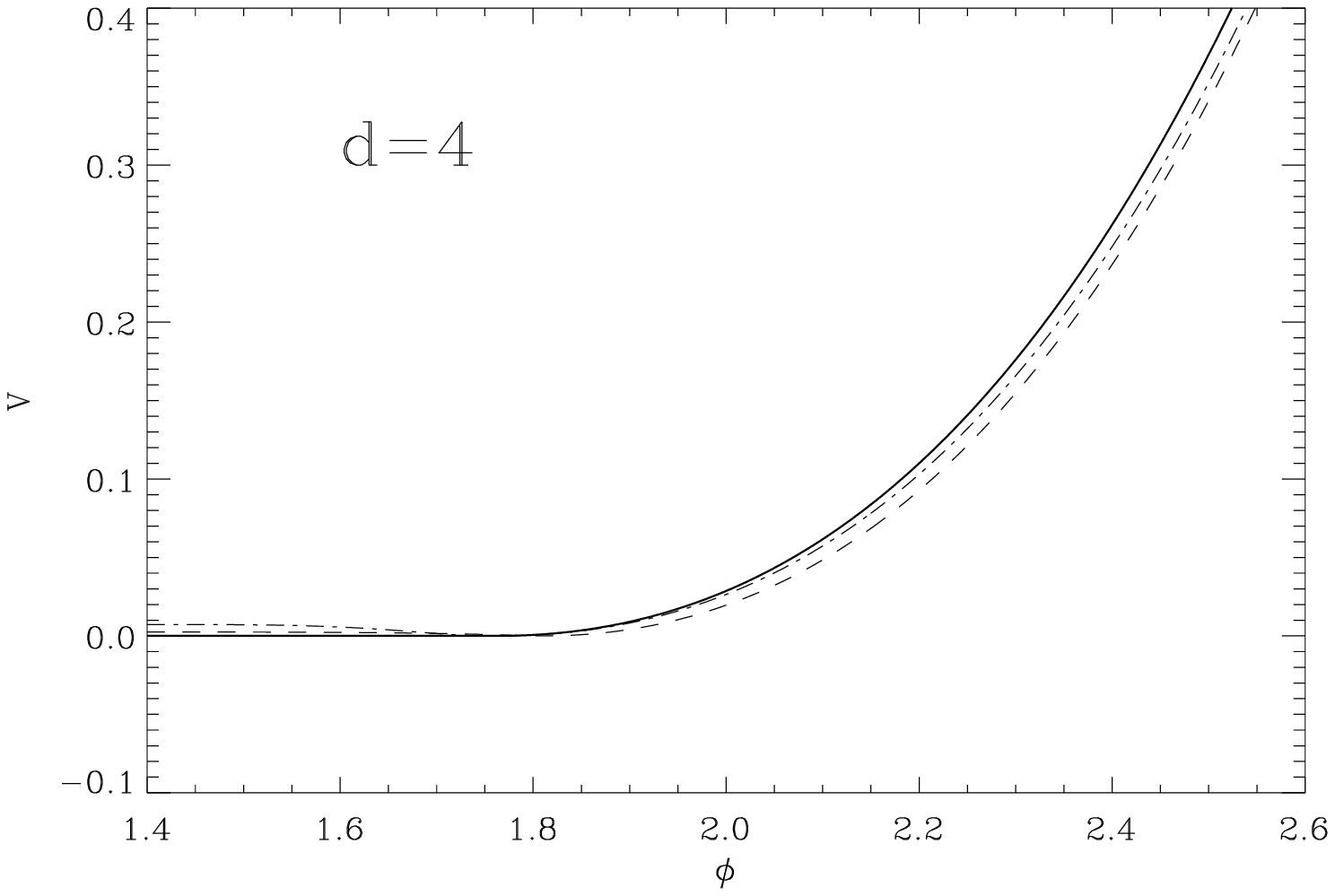}}
\caption{The blocked potential in $d=4$. The solid line is for $n=2$ (WH), the 
dashed is for $n=3$ and the dot-dashed is for $n=5$.}
\label{plot2}
\end{figure}
\renewcommand{\baselinestretch}{1.5}
\small\normalsize
\renewcommand{\baselinestretch}{1}
\small\normalsize
\begin{figure}[ht]
        \epsfxsize=14cm
        \epsfysize=10cm
\centerline{\epsffile{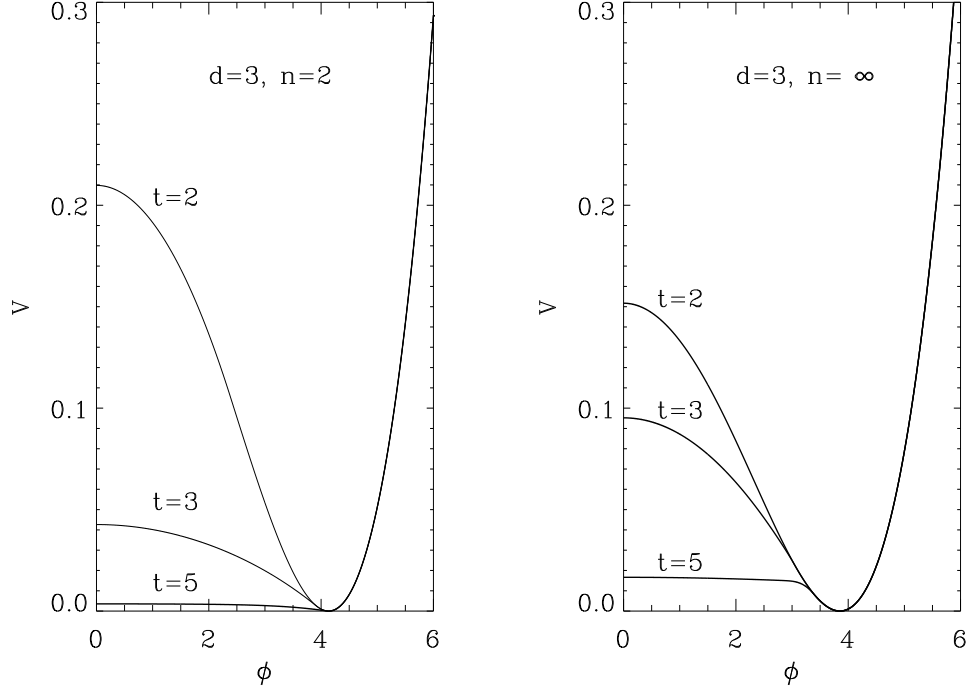}}
\caption{The blocked potential in $d=3$ for various values of the RG time $t$.}
\label{plot3}
\end{figure}
\renewcommand{\baselinestretch}{1.5}
\small\normalsize
\renewcommand{\baselinestretch}{1}
\small\normalsize
\begin{figure}[ht]
        \epsfxsize=14cm
        \epsfysize=10cm
\centerline{\epsffile{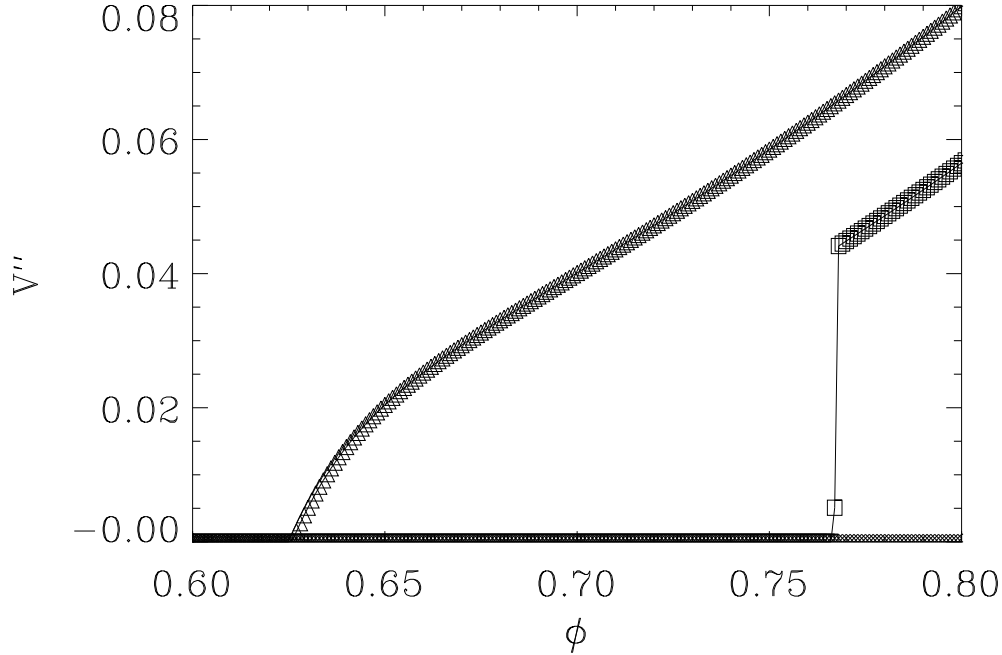}}
\caption{ $V''$ as a function of $\phi$ in the broken phase for $d=3$ and with $r_0=-0.1$ and 
$g_0=1.2$. The squares are for $n=3$ while the triangles 
are for the WH equation. The discontinuity is clearly visible in the first case, while it is
not present for the WH equation.}
\label{disc3}
\end{figure}
\renewcommand{\baselinestretch}{1.5}
\small\normalsize
\renewcommand{\baselinestretch}{1}
\small\normalsize
\begin{figure}[ht]
        \epsfxsize=14cm
        \epsfysize=10cm
\centerline{\epsffile{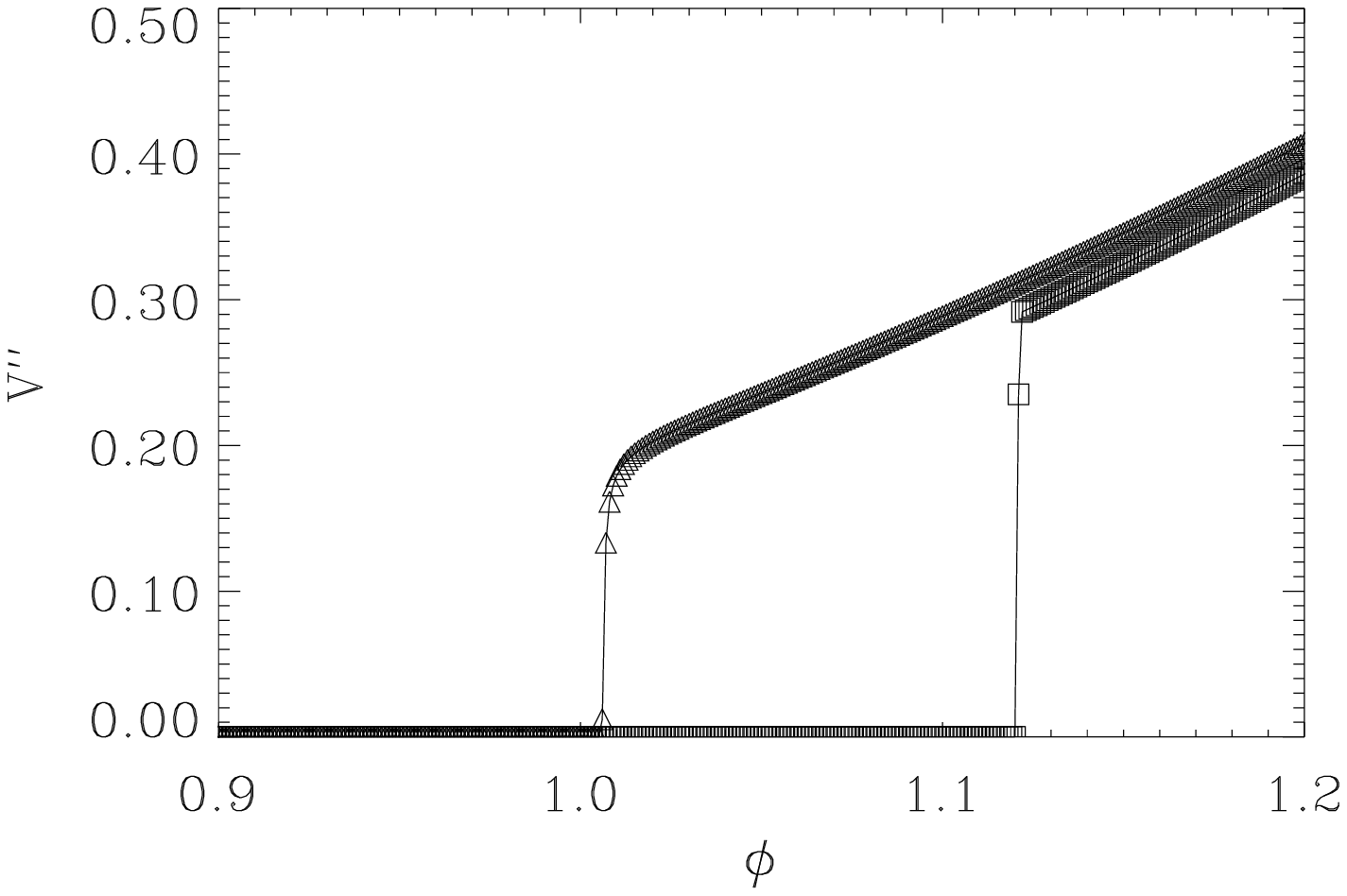}}
\caption{ $V''$ as a function of $\phi$ in the broken phase for $d=3.5$ and with $r_0=-0.1$ and 
$g_0=1.2$. The squares are for $n=3$ while the triangles 
are for the WH equation. The jump is clearly seen for the squares while there is a continuous
transition for the WH equation}
\label{disc3.5}
\end{figure}
\renewcommand{\baselinestretch}{1.5}
\small\normalsize

For the sake of completeness we notice that 
the $n\to \infty$ limit of (\ref{eq01}) reads
\be\label{eq02}
W''_k = 2 k^{-d+2} \ln W_k   
- \frac{k^{-d+2}}{W_k}\;
k {\partial W_k \over \partial k}
\ee
while if we instead define $W_k(\phi) = \ln(1+V''_k/k^2)$,   Eq.(\ref{setI}) implies 
\be\label{eq03}
W''_k 
= 4 k^{-d+2}  - \;
 2 e^{W_k} \; k {\partial W_k \over \partial k}.
\ee
Although no simple analytical solution is available in the broken phase near $\phi=0$ 
for Eq.(\ref{eq02}) we can still understand how the regulator affects
the approach to the spinoidal line. 

In fact, when $W_k$ is large and negative in Eq.(\ref{eq03}),
the diffusive term $\partial W/\partial k$ is exponentially
suppressed in this case and we obtain Eq.(\ref{inner1}) with $n=1$ (see also \cite{parola}).
On the contrary when $W_k$ is large and positive in Eq.(\ref{eq01}) and Eq.(\ref{eq02}) 
the diffusive term is only power-law suppressed. In particular the suppression is the 
slowest for $n \to \infty$,  like $1/W_k$, while if $n=d/2+\epsilon$ 
(being $\epsilon$ small and positive), the 
suppression is like $W_k^{-(1+\epsilon)/\epsilon}$ and 
(\ref{u2}) is clearly reached faster as $\epsilon$ gets smaller.
In other words convexity is best achieved with a small value of $\epsilon$.

An important issue related with the above discussion 
is to understand how the inner solution joins the 
outer region $\phi>\phi_0$.  For Eq.(\ref{setI}) it is known that 
a fixed point solution for $k\to 0$ is present, 
but it predicts a continuous second derivative of the effective potential
for $d<4$, which corresponds to a diverging compressibility at the transition
(see \cite{parola} for an extended discussion on this point).
The relevant question is whether the use of a smooth cutoff regulator 
as Eq.(\ref{setII}) and Eq.(\ref{setIV}) can cure this pathological behavior
of the ``exact'' WH equation. 

In order to investigate these problems in detail one must handle the
problem numerically, by extracting an accurate numerical solution of
the flow equation.  We thus have solved Eq.(\ref{eq01}) and
Eq.(\ref{eq02}) with the fully-implicit, predictor corrector
finite-difference scheme described in \cite{ames} for which a
rigorous result  ensures convergence
to the real solution for our numerical discretization grid.

Let us then write the bare potential as $V_\Lambda(\phi) = {r_0}\phi^2/2 
+ {g_0}\phi^4 /4!$, and $|r_0|<1$ being the bare mass measured in cutoff units.
Fig.(\ref{plot1}) and Fig.(\ref{plot2}) 
show the blocked potential for $d=3$, and $d=4$, respectively, in the broken phase.
As expected, if we push the integration closer to the $k\to 0$ limit all
the curves approach a completely flat bottom as it is shown in
Fig.(\ref{plot1}) and Fig.(\ref{plot2}) in $d=3$ and $d=4$, respectively,
for various values of $n$. In this case $r_0=-0.6$, 
$g_0=1.0$, and the final value of the RG ``time'' is $t=-10$. 
Different values of $r_0$ and $g_0$ leads to qualitatively similar results in the sense that,
as long as we are below the critical line, the flow always approaches the 
flat bottom convex potential. 

As we discussed above, we also notice from Fig.(\ref{plot2}) 
that the convergence to convexity is much faster for smaller values
of $n$. A plot for different values of the RG time $t$ is 
depicted in Fig.(\ref{plot3}) where the expected behavior is discussed: 
as it is apparent from these plots the the potential for $n=2$ is much
closer to the flat and convex solution already for $t=5$, than the $n=\infty$
case where, although a flat bottom is present near $\phi=0$, 
convexity is not achieved yet.
It is numerically difficult to reach higher values of 
$t$ while keeping the mesh spacing constant if $n \to \infty$. We find that in order to
reproduce the convexity at an acceptable level 
($|min(U)-U(0)| \sim 10^{-2}$ for $g_0= O(1)$)
the time step and the mesh spacing have to be both of the order of at least $10^{-3}$, 
which is not very efficient from the numerical point of view.  
On the contrary, if we choose $n=d/2+\epsilon$ ($\epsilon$ positive and small), then 
$|min(U)-U(0)|\sim 10^{-3}$ already with a mesh spacing 
one order of magnitude greater. Although we find numerical evidence that 
for $n\to \infty$ the solution approaches a flat bottom near the origin, 
we cannot exclude that convexity is never reached in this case.

An important result of our analysis is that the second
derivative of the potential shows the expected discontinuity for
$3\leq d<4$, for any $n>d/2$, as opposed to the ``exact'' WH equation, for
which the spinoidal line merges with the coexistence line in $d=3$.
This is clearly shown in Fig.(\ref{disc3}), where the discontinuity is visible
in the numerical output because always only one grid point is present
in the jump, and this feature does not depend on further refinements
of the spatial grid. Similar behavior is also observed in $d=3.5$, as 
it is apparent from Fig.(\ref{disc3.5}). 
\renewcommand{\baselinestretch}{1}
\small\normalsize
\begin{figure}[ht]
        \epsfxsize=8cm
        \epsfysize=8cm
\centerline{\epsffile{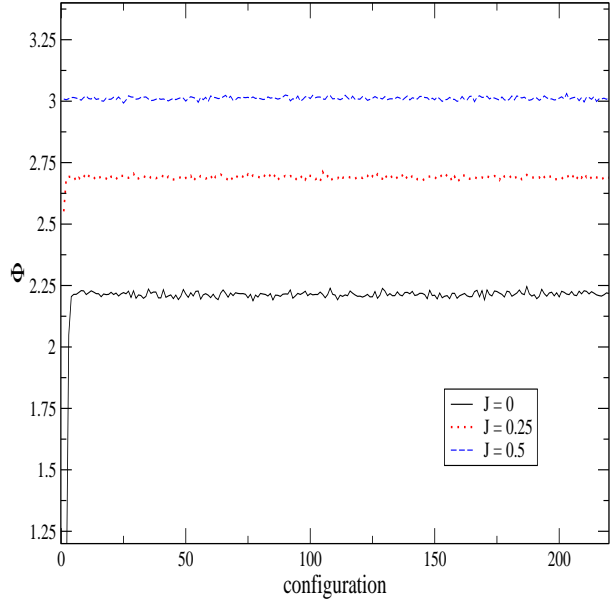}}
\caption{The average field as evaluated by the HMC simulation, in the $d=3$ case ($L=16$), 
with parameters $a^2r_0=-0.2, a g_0=0.24$. 
The results for three different values of $J$ are shown. }
\label{plot_mc_1}
\end{figure}
\renewcommand{\baselinestretch}{1.5} \small\normalsize
\renewcommand{\baselinestretch}{1}
\small\normalsize
\begin{figure}[ht]
        \epsfxsize=14cm
        \epsfysize=10cm
\centerline{\epsffile{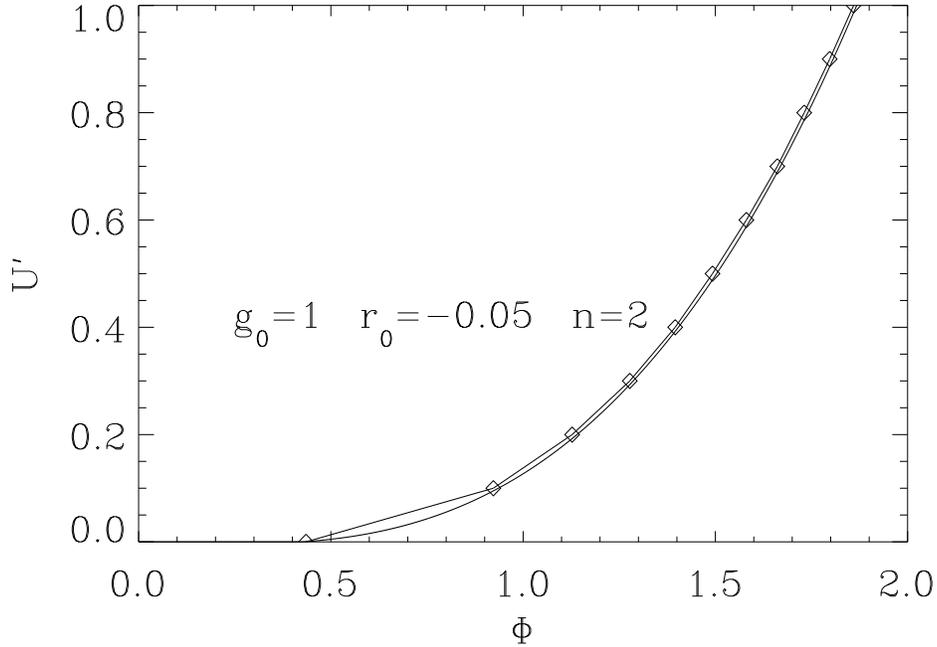}}
\caption{The blocked potential from the flow equation in $d=3$ 
vs the MC simulation for a volume of $12^3$ in the weak coupling case (MC errors negligible on this scale).}
\label{mc2}
\end{figure}
\renewcommand{\baselinestretch}{1.5}
\small\normalsize
\renewcommand{\baselinestretch}{1}
\small\normalsize
\begin{figure}[ht]
        \epsfxsize=14cm
        \epsfysize=10cm
\centerline{\epsffile{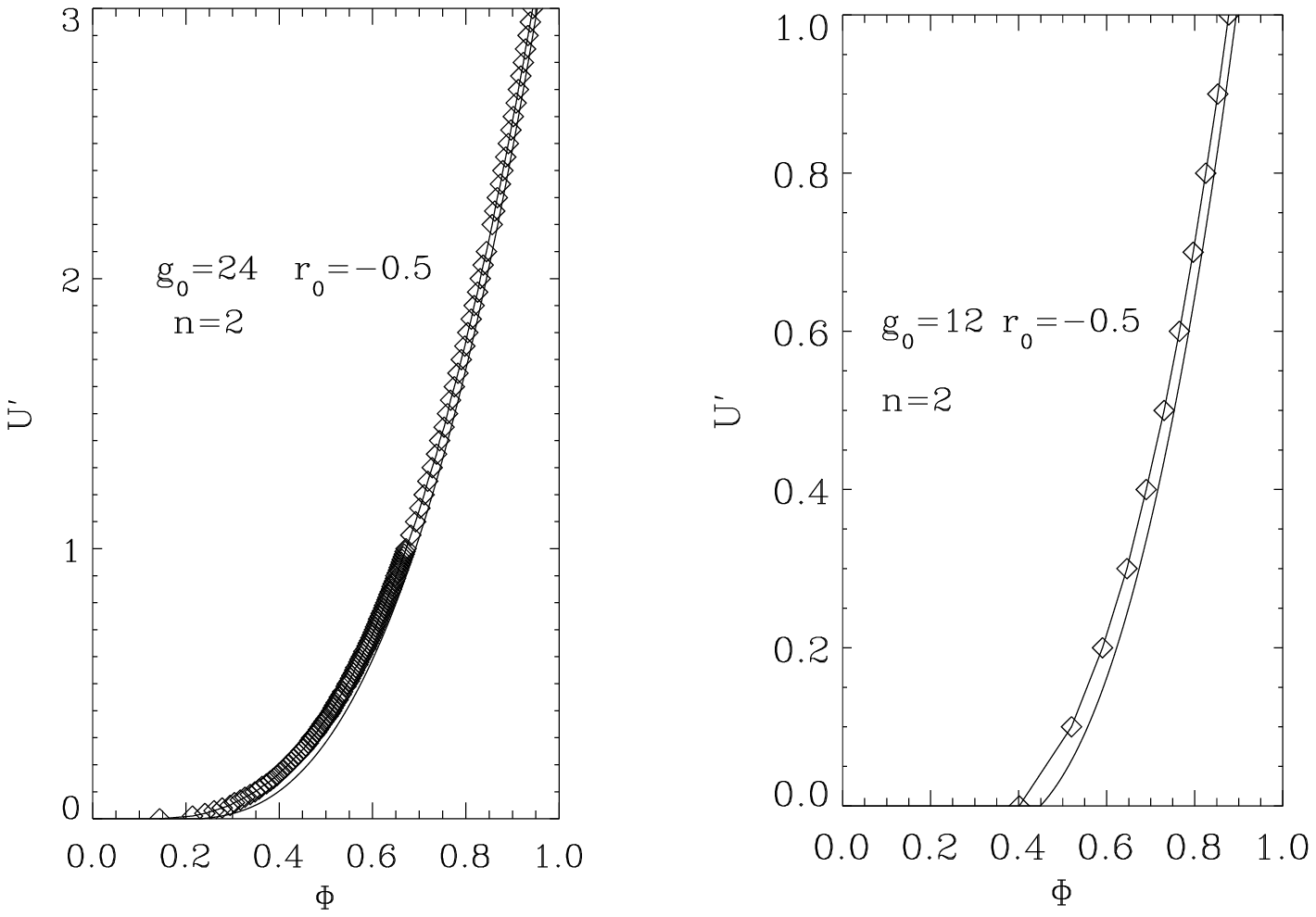}}
\caption{The blocked potential from the flow equation in $d=3$ 
vs the MC simulation for a volume of $12^3$ in the strong coupling case 
(MC error bars invisible on this scale).}
\label{mc1}
\end{figure}
\renewcommand{\baselinestretch}{1.5}
\small\normalsize

\section{Monte Carlo simulations}
\renewcommand{\theequation}{4.\arabic{equation}}
\setcounter{equation}{0}
\label{4S2}

Monte Carlo simulations of the $\phi^4$ model have been performed on
$d=2,3,4$ lattices in order to evaluate the Effective Potential and
compare its determination with the numerical solution of the flow
equations. In a lattice simulation, the inverse lattice spacing acts
as an UV momentum cutoff, while the finite size introduces an IR
cutoff. The model was discretised following the standard approach in
\cite{montvay}, and the field configurations were generated with the
Hybrid Monte Carlo algorithm. Introducing an uniform external current
$J$, a discrete version of the action can be written as follows:

\begin{equation}\label{laction}
S[\phi] = a^d \sum_n \left[ \frac{1}{2a^2}
\sum_{\mu=1}^d(\phi_{n+\mu}-\phi_n)^2+V(\phi_n)-J\phi_n \right]
\end{equation}

where $n$ is the site index, $a$ is the lattice spacing and

\begin{equation}\label{lpo}
V(\phi) = a^2\frac{r_0}{2}\phi^2+a^{4-d}\frac{g_0}{4!}\phi^4
\end{equation}

Introducing the normalization $a^{d/2-1}\phi \to (2\kappa)^{1/2}\phi$
and $a^{d/2+1}(2\kappa)^{1/2}J\to J$, the discretised, dimensionless action is:
\begin{equation}
S[\phi] = \sum_n \left[ -2\kappa \sum_{\mu=1}^d \phi_{n+\mu}\phi_n  + \phi_n^2 + 
\lambda(\phi_n^2-1)^2 - \lambda - J\phi_n\right] 
\end{equation}
where
\be
a^2r_0 =  \frac{1-2\lambda}{\kappa}-2d \;\;\;\;a^{4-d}g_0  =  \frac{6\lambda}{\kappa^2}
\ee

Periodic boundary conditions have been assumed. Both cold and hot
starts have been tested, and once thermalization was achieved, the
average of the field over the entire lattice was computed on each
configuration:

\begin{equation}
{\Phi}=\frac{1}{L^d}\sum_n \phi_n
\end{equation}

where $L$ is the number of lattice sites in each direction. The
results were saved every $50$ configurations to decorrelate the
results. In each case, $200$ uncorrelated configurations were
collected in order to get a statistically accurate determination of
the average field. It should be noticed that the size of the
statistical fluctuations decreased as the external current $J$ was
increased, as expected. In all cases, statistical errors have been
estimated using the bootstrap technique ($1000$ samples) and found to
be negligible.  Simulations have been run with different lattice
volumes, and finite size effects have been found to be negligible. For
each choice of the numerical parameters $(\kappa,\lambda)$, the simulation was
run with several values of the external current $J$.  This procedure
was not very different from that sketched in \cite{ardekani}. The
relation ${\Phi}(J)$ was then inverted to get $J(\Phi)$, and thus the first
derivative of the Effective Potential by
\begin{equation}\label{up}
J = U'(\Phi)
\end{equation}

We would like to compare $U'$ as computed from (\ref{up}) with the
same quantity obtained by numerical integration of the flow equation.
Although it would be possible to compare the constraint effective
potential \cite{oraife} at a given scale $k\propto 1/\Omega$, with the blocked
potential as computed from the flow equation at a scale $k$, it is
more interesting to focus on the $k\to 0$ limit, where possible
non-universal features due to the regulator may disappear. As we
discussed in the introduction we prefer to use the external current
method because it is simpler to implement and as accurate as the
constraint effective potential method \cite{ardekani}.

We explore a set of parameters which is not close to the critical
line, so that finite-size effects can be neglected, and we are able to
compare the result of the flow equation directly to the lattice
determination of the effective potential. In particular we find that
it is was not necessary to construct a lattice version of the RG flow
equation as discussed in \cite{shepard}. We have integrated
Eq.(\ref{eq01}) rewriting all the relevant quantities in units of the
UV cut-off and then we have followed the evolution down to $t\to \infty$ with
the help of the numerical integrator. In order to show the predictive
power of our approach we have not fine-tuned the bare parameters in
the RG flow equations to reproduce renormalized mass and coupling
constant obtained in the lattice calculation as done in \cite{shepard}.
Instead, we have decided to set the same bare mass and coupling
constant in the lattice bare potential (\ref{lpo}) and in the bare
potential of the RG equation. Moreover we have rescaled back our potential
and field according to (\ref{sca}) so that we get $U'_{k=0}(\Phi)$ out of 
the numerical computation. According to the analysis of the previous session
we have considered $n \in (d/2,d] $ where the numerical stability is best achieved. 

The results  are shown in Fig.(\ref{mc2})
and Fig.(\ref{mc1}) for $n=2$ and $d=3$, where it is apparent that there is already a very good 
agreement with the MC data both in the weak and in the strong coupling regime.
Better agreement could probably  be achieved by including the 
wave-function renormalization function, but this is not our main concern in this investigation. 

\section{Conclusions}
We have discussed the PTRG flow equation below the critical line in a scalar theory. 
In particular we have shown that the convexity property of the free energy is recovered 
by integration of the LPA flow equation in the $k\to \infty$ limit. Within a class
of $n$-dependent proper time regulator, the approach to the correct flat bottom potential 
is faster when $n = d/2 +\epsilon$ being $\epsilon$ positive and small. 
The expected discontinuity of the $U''$ at the transition is correctly reproduced for any 
value of $\epsilon >0$ as opposed to the ``exact'' WH flow ($\epsilon =0$), 
which does not show this feature. 
We have performed an extensive MC investigation of the EP in $d=3$ in order to discuss the 
numerical predictions and we found very good agreement both the strong coupling and weak coupling
phase without resorting to a fine-tuning procedure between the bare parameters in the MC and in the
RG flow equation.  We anticipate that our result can be relevant in gauge theory where the presence of 
the PT regulator is an essential tool deriving a non-perturbative flow equation \cite{liaoym}.

\section*{Acknowledgements}
We acknowledge Martin Reuter and Dario Zappal{\'a} for useful
comments. G. Lacagnina acknowledges the financial support by the
DFG-Forschergruppe ``Lattice Hadron Phenomenology'' and wishes to
thank V. Braun, A. Sch{\"a}fer and M. G{\"o}ckeler for useful discussions.

\end{document}